\documentclass[prd,notitlepage,nofootinbib,superscriptadress,12pt]{revtex4-1}

\usepackage{amsmath, amssymb}
\usepackage[]{graphicx}
\usepackage[caption=false]{subfig}
\usepackage{hyperref}
\usepackage{dsfont}
\usepackage{xcolor}
\usepackage{amsmath,amssymb,amsfonts, bm,bbm,slashed, subdepth}
\usepackage[normalem]{ulem}
\usepackage{hyperref}
\usepackage{cleveref}
\usepackage{enumerate}
\usepackage{setspace}
\usepackage{booktabs, tabularx}

\usepackage[paper=a4paper,
  includefoot, 
  marginparwidth=0.0mm, 
  marginparsep=1.3mm, 
  margin=20mm, 
  includemp]{geometry}

\newcommand{\be}{\begin{equation}}
\newcommand{\ee}{\end{equation}}
\newcommand{\ba}{\begin{array}}
\newcommand{\ea}{\end{array}}
\newcommand{\bea}{\begin{eqnarray}}
\newcommand{\eea}{\end{eqnarray}}
\newcommand{\balg}{\begin{align}}
\newcommand{\ealg}{\end{align}}
\newcommand{\bit}{\begin{itemize}}
\newcommand{\eit}{\end{itemize}}
\newcommand{\trm}[1]{\textrm{#1}}

\newcommand{\Mpc}{\trm{\Mpc}}
\newcommand{\yr}{\trm{\yr}}
\newcommand{\eV}{\trm{\eV}}

\begin{document}

\title{Attenuation effect and neutrino oscillation tomography}

\author{
 A. N. Ioannisian$^{1,2}$ and A. Yu. Smirnov$^{3}$
 }
\email{smirnov@mpi-hd.mpg.de}
\affiliation{
$^1$
Yerevan Physics Institute, Alikhanian Br.\ 2, 375036 Yerevan,
Armenia\\
$^2$Institute for Theoretical Physics and Modeling, 375036
Yerevan, Armenia\\
$^3$
Max-Planck-Institut f\"{u}r Kernphysik, Saupfercheckweg 1, D-69117 Heidelberg, Germany 
}

\begin{abstract}

Attenuation effect is the effect of  weakening of  
contributions to the  oscillation signal 
from remote structures of matter density profile.
The effect is a consequence of 
integration over the neutrino energy within the energy 
resolution interval. Structures of a density profile situated 
at distances larger than the attenuation length,  $\lambda_{att}$,  are not ``seen''.  
We show that the origins of attenuation are (i) averaging of oscillations in certain layer(s) 
of matter, (ii) smallness of matter effect:  $\epsilon \equiv 2EV/\Delta m^2 \ll 1$, 
where $V$ is the matter potential, 
 and (iii) specific initial and final states on neutrinos. 
We elaborate on the graphic description of the attenuation  which allows 
us to compute explicitly the effects in the $\epsilon^2$ order for 
various density profiles and oscillation channels.    
The attenuation in the case of partial averaging is described. 
The effect is crucial for interpretation of oscillation data and 
for the oscillation tomography of the Earth with low energy (solar, supernova, 
atmospheric, {\it  etc.}) neutrinos.

\end{abstract}

\pacs{14.60.Pq, 26.65.+t, 91.35.-x,95.85.Ry, 96.60.Jw, }

\maketitle

\section{Introduction
\label{sec:intro}}

The {\it attenuation effect} introduced in \cite{Ioannisian:2004jk} is 
the key element for understanding neutrino oscillations in the Earth.  
It describes  weakening of contribution of a remote structure of a 
matter density profile to the oscillation signal in a detector.    
The contribution decreases with increase of distance between a structure and a detector 
because of  finite accuracy of reconstruction of the neutrino energy. 
The latter can be due to 
finite energy resolution of a detector,  or finite width  of  produced neutrino  
energy spectrum,  or due to  kinematics of 
process when neutrino energy can not be obtained  uniquely.   
The better the energy resolution, the more remote structures can be observed. 
The attenuation effect applies to the  astrophysical  (solar, supernova)  neutrinos 
arriving at the surface of the 
Earth as incoherent fluxes of mass states. It also applies to the terrestrial neutrinos 
of different origins: the reactor antineutrinos, atmospheric neutrinos of low energies,  {\it etc}.   

The attenuation effect explains why, e.g., 
Super-Kamiokande can not observe the core of the Earth using the solar neutrinos.
The computed curves (see fig.~1 in \cite{Renshaw:2013dzu}) 
do not show increase of the $\nu_e$ regeneration for the zenith angles $|\cos \theta_z| > 0.83$ 
(core crossing trajectories) in spite of 2 times larger 
density in the core than in the mantle.  
The zenith angle dependence of rate of events is flat as it would be in the absence of the core.  
In the case of Super-Kamiokande, which detects the boron neutrinos with continuous 
energy spectrum,   the attenuation is due to integration over the energy of neutrino  
since the observed signal is the recoil electron from the $\nu -e$ scattering. 
As can be seen in fig.~53 of  \cite{Hosaka:2005um} and  fig.~2 of \cite{Smy:2003jf},  
even selection of narrow intervals for the recoil electron energy 
does not improve the sensitivity to the core. 
One can observe  only small spikes at $|\cos \theta_z| = 0.83$ 
which are about 30 times smaller than the difference of the night and day signals. 
The spikes slightly increse in the  high recoil energy bins 
$(15 - 16)$ MeV and $(16 - 20)$ MeV. The reason is that only very high energy 
part of the neutrino spectrum contributes to these bins. Selecting these bins 
we are effectively narrowing the integration interval over neutrino energies. 
In contrast,  small density jumps close to the surface of the Earth 
($|\cos \theta_z| \sim  0.05 - 0.2$) produce much stronger effect. 

The core can be seen, in principle,  using the beryllium neutrinos 
with relative width of the line $\sigma_E/E \sim 0.005$ \cite{Ioannisian:2015qwa}. 
Good energy reconstruction can be achieved in experiments based on 
the $\nu-$nuclei scattering. As an example,  the $\nu -$ Ar interactions  have been 
considered \cite{Ioannisian:2017dkx} and the energy-nadir angle distribution of events 
during the night time has  been computed. 
It is shown that  the nadir angle dependence of the night excess can be interpreted as 
hierarchical perturbations of the result 
for constant density profile. 
The strongest perturbation of the lowest order is produced by the closest to a detector 
density jumps in the outer mantle. The first order dependence is then perturbed by 
smaller size effect of the deeper mantle jumps. In turn, this second order approximation is  
perturbed by the Earth core effect. 
With improvement of energy resolution of a detector the effect of remote structures increases 
~\cite{Ioannisian:2017dkx}. 
The core can be seen with  $\sigma_E/E < 0.1$.  

The attenuation effect has been obtained for 
the mass-to-flavor transition, e.g.  $\nu_1 \rightarrow \nu_e$. 
Paradoxical result is that for the inverse channel,  the 
flavor-to-mass transition $\nu_e \rightarrow \nu_1$, 
the situation is opposite:    
close to a detector structures are attenuated,  whereas 
remote structures can be seen. 
In the $\nu_e \rightarrow \nu_e$ channel the attenuation may or may not be realized  
depending on features of the profile. 

In this paper we further elaborate on physics of the attenuation effect. 
We clarify the meaning of the attenuation length. Using simple examples we 
(i) formulate conditions for realization of the effect, 
(ii) show its specific properties  for various density profiles and channels, 
(iii) prove that it is realized in the lowest order in $\epsilon$,  
(iv) consider the cases of partial attenuation. 
The geometric (graphic) description of the attenuation effect  
allows us to understand the paradoxical features described above. 
We show that the attenuation effect is 
a result of certain averaging of oscillations,  smallness 
of mixing of the neutrino mass states in matter and specific initial and final states of 
neutrinos. Using the graphic representation we compute effects 
in all  orders in $\epsilon$. 
We consider effects in  pure flavor channels 
which can be applied to  neutrinos of terrestrial origins:   
reactor antineutrinos,  low energy atmospheric neutrinos, geoneutrinos, 
neutrinos from $\pi-$ and $\mu-$ decay at rest.  

The results obtained here are important for interpretation of data on neutrino 
oscillation in the Earth and for planning of future experiments aimed at the 
neutrino oscillation tomography. 

The paper is organized as follows: In Sec.~\ref{sec:attenuation} we recall 
the main points  of derivation of the attenuation effect. In Sec.~\ref{sec:decoh} we clarify  
meaning of the attenuation length and present 
graphic description of the effect. 
The effect in two layers of matter is discussed in Sec.~\ref{sec:twolayers}.  
The case of multi-layer medium is explored in 
Sec.~\ref{sec:threelayers}. In Sec.~\ref{sec:partial} 
we consider the case of partial averaging. Discussion and 
conclusions  are presented in Sec.~\ref{sec:discussion}.

\section{Attenuation effect
\label{sec:attenuation}}

Astrophysical (solar, supernova)  neutrinos 
arrive at the Earth as incoherent  fluxes of the mass eigenstates $\nu_i$. 
In the Earth, each mass state ``splits" into eigenstates in matter 
and oscillates. The mixing angle of the $\nu_1$ and $\nu_2$ 
mass states in matter, $\theta'$,  is determined by 
\be
\sin 2 \theta' = 
\frac{c_{13}^2 \epsilon \sin 2\theta_{21}}{\sqrt{(\cos 2
\theta_{12} - c_{13}^2\epsilon)^2 + \sin^2 2 \theta_{12}}} 
= c_{13}^2 \epsilon \sin 2\theta_{21}^m
\approx \epsilon \sin 2\theta_{21}. 
\label{eq:massangle}
\ee
Here $\theta_{21}^m$ is the  flavor mixing angle in matter,  $c_{13} \equiv \cos \theta_{13}$,  and 
\be
\epsilon   \equiv  \frac{2 V_e E}{\Delta m_{21}^2 } = 0.03 \left(\frac{E}{10 {\rm{MeV}}}\right) 
\left(\frac{\rho}{2.6 {\rm g/cm^3}}\right).      
\label{eq:epsil}
\ee
Thus, the mixing angle of the mass states in matter 
is suppressed by $\epsilon$.  In the Earth for $E< 30$ MeV  (solar,  supernova  neutrinos) 
oscillations proceed  in the low density regime when $\epsilon \ll 1$. 
The splitting of the eigenvalues of Hamiltonian:  
\begin{equation}
\Delta_{21}^m  \equiv \frac{\Delta m^2_{21}}{2E} \sqrt{(\cos 2
\theta_{12} - c_{13}^2\epsilon )^2 + \sin^2 2 \theta_{12}} \ 
\label{split}
\end{equation}
determines the oscillation length in matter  
\begin{equation}
l_m = \frac{2\pi}{\Delta^m_{21}} = l_\nu [1 + 
\cos 2\theta_{12} c_{13}^2 \epsilon + O(\epsilon^2)], 
\label{osclength}
\end{equation}
which is  close to the vacuum oscillation length $l_\nu = 4\pi E/ \Delta m^2_{21}$.

Detector registers the flavor states and for definiteness we will take $\nu_e$. 
Therefore the relevant transition is $\nu_i \rightarrow \nu_e$. Furthermore, 
at low energies, when matter effect on the 1-3 mixing is negligible, 
it is enough to find the transition for one mass state, 
and other can be obtained using unitarity.  
So,  in what follows for definiteness we will focus on the $\nu_1 \rightarrow \nu_e$ transition.

Without matter effect the probability equals  $P_{1e} =  |U_{e1}|^2$, 
where  $U_{e1}$ is the $e1-$element of the 
mixing matrix in vacuum. Therefore 
the Earth matter effect is given by the ``regeneration factor''
$$
f_{reg} \equiv P_{1e} - |U_{e1}|^2. 
$$ 
In the lowest order in $\epsilon$ the factor equals  
\cite{Ioannisian:2004jk},  \cite{Akhmedov:2004rq}
 \begin{equation}
  \label{p1} 
f_{reg} =
C \int_{0}^{L} \! \! \! \! dx \  V_e(x) \sin \phi^m_{x \to L} ,
\end{equation}
where $C \equiv  - {1 \over 2} \sin^2 2 \theta_{12} c_{13}^4$    
and 
\begin{equation}
\phi^m_{x \to L} (E) \equiv \int_{x}^{L} \! \! d x \ \Delta_{21}^m(x, E) 
\label{phasexl}
\end{equation}
is the phase acquired from a given point of trajectory $x$  to a detector.  
$L$
is the total length of trajectory.  

The attenuation effect is a consequence  of  integration of the 
oscillation probability over the neutrino 
energy with the neutrino energy reconstruction  function   
$g(E_r, E)$, where $E_r$  and $E$ are the reconstructed  and true energies 
correspondingly.  The  width  of $g(E_r, E)$ is determined by the smallest quantity among 
(i) a width of neutrino spectrum, (ii) energy resolution of a detector, 
(iii) an accuracy of the neutrino energy reconstruction determined by kinematics of the process 
used for a detection. The regeneration factor 
averaged over the energy equals
\begin{equation}
\bar{f}_{reg}(E_r) = \int dE~ g(E_r, E) f_{reg} (E, x) \ ,  
\label{avv}
\end{equation}
with  $\int dE g(E_r, E) = 1$.  Inserting (\ref{p1}) into  (\ref{avv}) we obtain 
\begin{equation}
\bar{f}_{reg}(E_r)  =  C
\int_{0}^{L} \!   dx \ V(x) { F(L-x)} \sin \phi^m_{x \to  L}(E_r),
\label{eq:attenuation}
\end{equation}
where $F(d)$  is the attenuation factor \cite{Ioannisian:2004jk} 
and  $d \equiv  L - x$ 
is the distance from a given structure of the profile to a  detector. 
The factor $F$ is defined by the equality
\begin{equation} 
F(L-x) \sin \phi^m_{x \to  L}(E_r) = \int dE~ g(E_r, E) \sin \phi^m_{x \to  L}(E)
\label{eq:att-fact}
\end{equation}
in such a way that for the ideal energy resolution, $g(E_r, E) = \delta(E -  E_r)$, 
one would get $F(L-x) = 1$, {\it i.e}, attenuation is absent. Notice that 
due to presence of sine of the phase in the integral (\ref{eq:att-fact}) 
the factor $F(L-x)$ appears as  a kind of Fourier transform of the energy 
resolution function $g(E_r, E)$.  

For the Gaussian  resolution function with width $\sigma_E$,  
\begin{equation}
 g(E_r, E)={1 \over \sigma_E \sqrt{2 \pi}}
e^{-{(E_r -E)^2 \over 2 \sigma_E^2}}, 
\label{eq:gaussian}
\end{equation}
we obtain from (\ref{eq:att-fact}) 
$$
F(d)\simeq e^{-2\left({ d \over \lambda_{att} }\right)^2},  
$$
where 
\begin{equation}
\lambda_{att} \equiv l_\nu \frac{E}{\pi \sigma_E}
\label{eq:attlength}
\end{equation}
is the {\it attenuation length}. If  $d = \lambda_{att}$,  
the factor equals $F(\lambda_{att}) = 0.135$,  and therefore according to   
(\ref{eq:attenuation}), a contribution to the oscillation effect 
of structures with $d > \lambda_{att}$  is strongly suppressed. 
As follows from (\ref{eq:attlength}),  the better the  energy 
resolution of a  detector, 
the more remote structures can be ``seen''. 
Thus, for the relative energy resolution $\sigma_E / E = 0.1$ and $l_\nu = 400$ km the attenuation 
length equals  1470 km 
and structures of a  density profile at $d > 1470$ km can not be observed. 
For  $\sigma_E / E = 0.2$ already structures with  $d > 750 $ km 
are strongly attenuated. 

The origin of the attenuation can be traced from  Eq. (\ref{eq:attenuation}) 
where the potential $V(x)$ is integrated with the sine of the phase  acquired  from 
coordinate of a structure, $x$,  to a detector.

Notice that computing number of events in a detector  we integrate 
over energy not just $f_{reg}$, as in (\ref{avv}), 
but the product of  $f_{reg}$ with    
the flux $F$ and cross-section $\sigma$. 
The  product $\sigma F$  depends on energy, but   
even in this case the results are qualitatively 
unchanged. 
If the energy resolution is high,  dependence on energy of  the product  $\sigma F$  
can be neglected and the product  can be 
put out of the integral.

The attenuation effect  is also realized  for  the flavor neutrinos of the 
terrestrial origins (low energy atmospheric neutrinos, neutrinos from 
pion and muon decay at rest). For these neutrinos loss of coherence 
can occur in the first layer of matter, e.g., the mantle of the Earth, 
so that at internal structures 
the incoherent flux of the eigenstates (close to mass states) arrives. 
Attenuation is then realized for inner structures.

In what follows we will consider mainly attenuation for   
the 1-2 mode of oscillations described by  
the 1-2 sub-system   of the complete $3\nu-$system.  
At low energies the third mass state, $\nu_3$,  decouples and dynamics of the 
$3\nu$ evolution  is reduced to the $2\nu$ evolution in the so called propagation basis 
(see \cite{Maltoni:2015kca} for details), which is related to the flavor basis, in particular, by the 
1-3 rotation on the angle $-\theta_{13}$. At this rotation $\nu_e \rightarrow \nu_e'$.  
The matter effect on the 1-3 mixing can be neglected and the remaining    
$2\nu-$sub-system is characterized by $\theta_{12}$, 
$\Delta m^2_{21}$ and the potential $c_{13}^2 V_e$. 
So,  in what  follows we will consider the $2\nu-$transition $\nu_1 \rightarrow \nu_e'$.  
(We will omit prime keeping in mind that  results in the flavor basis can be 
obtained from the results in the propagation basis by multiplying them by $c_{13}^2$.)

With this, the evolution in the Earth is reduced to the 
$2\nu$ evolution in the potential $V = c_{13}^2 V_e$. For simplicity  
we omit the subscript $\theta_{12} \rightarrow \theta$.  
Then using eq. (\ref{eq:massangle}) we find for the mixing of mass states
in matter 
\be 
\sin^2 \theta' \approx \frac{\epsilon^2}{4}
\frac{
\sin^2 2\theta}{(\cos 2 \theta - \epsilon )^2 + \sin^2 2 \theta} 
= \frac{1}{4} \epsilon^2 \sin^2 2\theta_m
\approx \frac{1}{4} \epsilon^2 \sin^2 2\theta .   
\label{eq:massangle1}
\ee 
Here  $c_{13}^2$ is included in the potential and $\epsilon$.
We comment on attenuation for the 1-3 mode  in Sec. \ref{sec:discussion}.

\section{Attenuation effect and decoherence 
\label{sec:decoh}}

Let us first clarify  the meaning of the attenuation length $\lambda_{att}$. 
According to (\ref{eq:attlength}) 
the phase acquired by neutrino with energy $E$ over the distance $\lambda_{att}$ equals 
$$
\phi(E) = 2\pi \frac{\lambda_{att}}{l_m} \approx 2\pi \frac{E}{\pi \sigma_E}.   
$$
Then the  difference of  phases of neutrinos with the difference of energies
$\Delta E$ is 
\be
\Delta \phi =  2\pi \frac{\Delta E}{\pi \sigma_E}.
\label{eq:f-diff}
\ee
For the integration interval,  
$\Delta E = \pi \sigma_E$ the  
eq. (\ref{eq:f-diff}) gives $\Delta \phi =  2\pi$. 
Therefore integration over the energy resolution  leads to averaging of oscillations.  
So,  $\lambda_{att}$ is the distance (or width of the layer) over which oscillations 
observed with the energy resolution $\sigma_E$ are averaged. \\

Let us consider a density profile with some structure, the $s-$layer, e.g. density bump 
at $x = 0 \div x_s$ and the ``decoherence'' layer $d$  at $x_s \div (x_s + x_d)$. 
The bump should have  sharp edges, so that the adiabaticity is broken.  
The decoherence layer has constant or slowly changing density.  
Suppose a neutrino enters the profile at $x = 0$, 
while  a detector is placed at $x = x_d + x_s$. 
The distance between the structure and  a detector equals $x_d$. 
(Actually the presence of  matter in the $d$ layer is not important.)  
The densities in $d$ and $s$ are low being  of the same order. 
Recall that the Earth density can be considered as layers with 
slowly changing density inside the layers and sharp density change 
on the borders between them \cite{Dziewonski:1981xy}. 
So, our consideration can be immediately applied to this realistic situation.

Suppose $d > \lambda_{att}$, so that oscillations in the 
$d-$layer are averaged (or equivalently, coherence of the 
neutrino state is lost). Let 
$\nu_{1m}^d$, $\nu_{2m}^d$ be the neutrino eigenstates in $d$. 
Suppose the mass state $\nu_1$ arrives at the $s-$layer and after oscillations 
in  $s$ enters the $d-$layer as $\nu_x$ which can be parametrized as 
\be
\nu_x = \cos \theta_x \nu_{1m}^d + \sin \theta_x \nu_{2m}^d e^{-i \phi_x}.  
\label{eq:nux}
\ee
So,  the information about the $s-$layer is contained in the 
angle $\theta_x$  and the phase $\phi_x$. 
It may happen that some averaging occurred already before arriving at $d$. 
This can be accounted by the overall normalization factor of $\nu_x$,  $N$,  
such that $|N|^2 < 1$. 
We assume also that 
\be
\theta_x = B \epsilon, ~~~~B  = {\cal O}(1). 
\label{eq:thetax}
\ee
and $\epsilon$ is defined in (\ref{eq:epsil}).  
The phase $\phi_x$ becomes irrelevant due to 
averaging  in the layer $d$.  

In terms of the eigenstates $\nu_{im}^d$ $(i = 1, 2)$ 
the electron neutrino  and the mass state $\nu_1$ 
are given by  
\be
\nu_e = \cos \theta_d~\nu_{1m}^d + \sin \theta_d~\nu_{2m}^d, ~~~~
\nu_1 = \cos \theta_d'~\nu_{1m}^d + \sin \theta_d'~\nu_{2m}^d, 
\label{eq:mixingsf1}
\ee
where $\theta_d$ and $\theta_d'$ are the mixing angles 
of the flavor states and the mass states 
in the  $d-$layer correspondingly. 
Then according to (\ref{eq:nux}) the probability to observe $\nu_e$  
in a detector equals 
\be 
P_x = |\langle \nu_e| \nu_x \rangle|^2 = \cos^2 \theta_x \cos^2 \theta_d 
+ \sin^2 \theta_x \sin^2 \theta_d.  
\label{eq:prob-x}
\ee
So, after averaging  the information about the structure  
is encoded in $\theta_x$ only. 

In the absence of $s-$layer, the neutrino $\nu_1$ enters immediately 
the layer $d$ and propagates there. 
Then instead of (\ref{eq:prob-x}),  we obtain the probability to detect 
$\nu_e$:  
\be 
P_1 = |\langle \nu_e| \nu_1 \rangle|^2 = \cos^2 \theta_d' \cos^2 \theta_d 
+ \sin^2 \theta_d' \sin^2 \theta_d. 
\label{eq:prob-1}
\ee    
The difference of the probabilities in (\ref{eq:prob-x}) and   
(\ref{eq:prob-1}),  which is the measure of  effect of the $s-$layer,  
equals
\be 
\Delta P_e \equiv P_x  - P_1 =  (\sin^2 \theta'_d   - \sin^2 \theta_x) \cos 2\theta_d. 
\label{eq:prob-diff}
\ee    
Since density of the structure is of the order of density in layer $d$, 
we obtain using (\ref{eq:thetax}) and (\ref{eq:massangle1})
\be 
\Delta P_e = P_x  - P_1 \approx (B^2 - 1) \frac{1}{4}\epsilon^2  \sin^2 2\theta_d \cos 2\theta_d 
\approx (B^2 - 1) \frac{\epsilon^2}{4} \sin^2 2\theta \cos 2\theta.                             
\label{eq:prob-diff2}
\ee
The equalities corresponds to the low density case.
So,  the effect of structure  is absent in the first order in $\epsilon$, 
{\it i.e.} attenuated,  
in agreement with our previous consideration. 
Its effect  appears in the second order in small parameter $\epsilon$. 

Averaging eliminates information about the phase, 
and therefore removes interference,  so that small parameters appear being squared.\\

\begin{figure}[!]
\hspace{0.1cm}
\includegraphics[width=0.45\textwidth, height=0.45\textwidth]{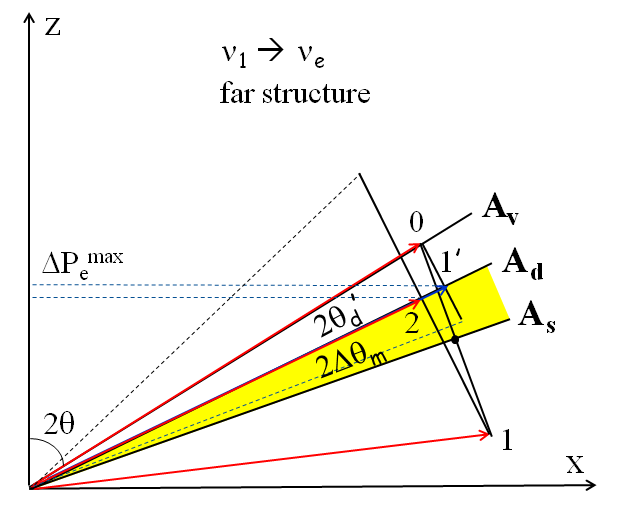}
\includegraphics[width=0.45\textwidth, height=0.45\textwidth]{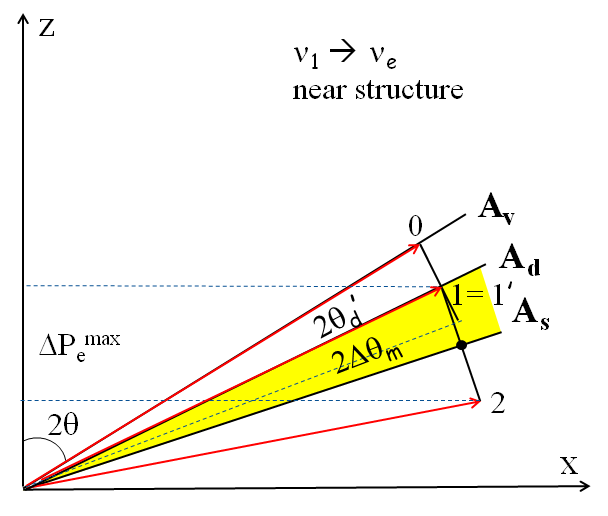}
\caption[�]{Graphic representation of evolution of the neutrino polarization vector 
in the case of  $\nu_1 \rightarrow \nu_e$
transition. Numbers indicate positions of the neutrino  
vector (red) at the  borders of different layers in the order of their crossing. 
They enumerate vectors  ${\bf P}_1$,  ${\bf P}_2$, ... (see the text). 
Position of the vector marked by ``0'' corresponds to the initial state. 
The positions of $1'$ marks the final state ${\bf P}^0$ (blue) in the absence  of the structure.    
To make effect visible we took rather large $\epsilon$.  
{\it The left panel:} remote structure.  
The diameter of precession $\sim \epsilon$ and its projection onto 
the eigenstate axis is given by $\epsilon$. This leads to the attenuation. 
{\it The right panel:}  the same as in the left panel but for near structure. 
The precession diameter equals $\sim \epsilon$, and its projection on the flavor 
axis is ${\cal O} (1)$. The attenuation is absent.
\label{fig:f1c}}
\end{figure}

This result as well as results for more complicated cases  
can be obtained easily  using the graphic representation of oscillations
based on  analogy of the oscillations and precession of the spin of electron 
in the magnetic field \cite{graphic} (see Fig.~\ref{fig:f1c} - \ref{fig:f8c}).  
According to this representation neutrino state is described by the polarization vector 
in the flavor space:  
$$
{\bf P} = \frac{1}{2} \bar{\psi} {\bf \sigma} \psi, ~~~~ \psi^T \equiv (\nu_e, ~ \nu_a). 
$$ 
Oscillations are equivalent to precession of the 
vector ${\bf P}$ in the flavor space 
$({\bf x}, {\bf y}, {\bf z})$ around the axis of eigenstates in matter ${\bf A}_m$. 
The axis lies in the $({\bf x}, {\bf z}) - $ plane 
and the  angle between the flavor axis ${\bf z}$  and ${\bf A}_m$ equals  
$2\theta_m$.  The direction of the axis of the mass states in vacuum,  ${\bf A}_v$,  
with respect to  ${\bf z}$ is given by the vacuum mixing angle $2\theta$.  
We use normalization  $|{\bf A}_m|^2 = |{\bf A}_v|^2 = 1$. 
The probability to observe $\nu_e$ is given by the projection of ${\bf P}$
onto the flavor axis ${\bf z}$: 
$$
P_e = ({\bf P}\cdot{\bf z} ) + \frac{1}{2}. 
$$

Let  ${\bf A}_d$ be the axis of eigenstates in the $d-$layer. 
Loss of coherence (averaging) in $d$ means that neutrino polarization 
vector ${\bf P}$ precesses around ${\bf A}_d$ with decrease of the orthogonal to ${\bf A}_d$ 
component. Projection of ${\bf P}$ on ${\bf A}_d$ does not change. 
Thus,  the vector ${\bf P}$ shrinks and eventually coincide  with 
its own projection onto ${\bf A}_d$: 
$$
{\bf P} \rightarrow ({\bf P} \cdot {\bf A}_d)~ {\bf A}_d . 
$$
Further on we will  consider attenuation effect in terms of this graphic representation. \\

Let ${\bf P}_x$ be the vector which describes the state $\nu_x$ in the example discussed above. 
The angle between ${\bf P}_x$ and the axis ${\bf A}_d$ equals $2 \theta_x$, and $\theta_x$ 
is defined in (\ref{eq:nux}). 
Averaging of oscillations in the layer $d$ means that 
${\bf P}_x$  evolves to its projection on ${\bf A}_d$:   
\be 
{\bf P}_x  \rightarrow  
\frac{1}{2} \cos 2\theta_x {\bf A}_d. 
\label{eq:xtoe}
\ee
Similarly,  the polarization vector ${\bf P}_1$, which corresponds to the mass state $\nu_1$, 
evolves (loosing the coherence) as 
\be 
{\bf P}_1 \rightarrow \frac{1}{2} \cos 2\theta'_d {\bf A}_d. 
\label{eq:1toe}
\ee
Here $2\theta'_d$ is the angle between ${\bf P}_1$ and ${\bf A}_d$  introduced 
in (\ref{eq:mixingsf1}).  Then the difference of the final vectors 
in (\ref{eq:xtoe})   and (\ref{eq:1toe}): 
\be
\frac{1}{2} (\cos 2\theta_x - \cos 2\theta'_d) {\bf A}_d. 
\ee
Projection of this difference onto the flavor axis ${\bf z}$  
(recall that $({\bf A}_d \cdot {\bf z}) = \cos 2\theta_d$)  presents effect 
of the $s-$layer on the $\nu_e$ survival probability:
$$
\Delta P_e  \equiv   P_x - P_1 = \frac{1}{2} (\cos 2\theta_x - \cos 2\theta'_d) \cos 2 \theta_d, 
$$
which coincides with expression in (\ref{eq:prob-diff}). \\

\section{Two layers case
\label{sec:twolayers}}

Let us consider oscillation effect in the $s-$layer explicitly,  
assuming first that the densities in 
$s$ and $d$ are constant.  
We denote by ${\bf A}_s$ the axis of eigenstates in the $s-$layer. 
A density jump on the border between the  $s-$ and $d-$layers leads to sudden 
change of the mixing angle in matter,  and  consequently, 
to change of direction of the  
eigenstate axis: ${\bf A}_s \rightarrow {\bf A}_d$. 
The angle between  ${\bf A}_s$ and ${\bf A}_d$ equals    
\be 
2 \Delta \theta_m \equiv 2\theta_s -  2\theta_d = 2\theta_s' - 2\theta_d'. 
\label{dmix}
\ee 
The parameter  
$$
J_m  \equiv \sin 2 \Delta \theta_m \sim \epsilon,  
$$
which we will call the jump factor, quantifies the effect of structure:  
the effect should be proportional to $J_m$, and  
if the structure is absent,  $\Delta \theta_m = 0$.  
Notice that $\Delta \theta_m$ is positive, if density in the $s-$layer 
is larger than that in the $d-$layer, and 
$\Delta \theta_m < 0$, if the density in $s$ is smaller. 
For definiteness we will present plots for $\Delta \theta_m > 0$. 
It is easy to see that formulas we will obtain are valid for both cases. 
If $\Delta \theta_m > 0$, the effect of structure 
on the $\nu_e$ probability is negative,  
$\Delta P_e < 0$. It is positive for $\Delta \theta_m <  0$.  
Oscillations in the $s-$layer may or may not be averaged. \\

To perform the   oscillation tomography 
(see e.g. 
\cite{Nicolaidis:1990jm, Lindner:2002wm, Akhmedov:2005yt, Winter:2006vg, Koike:2016jrb}) 
one can scan a  density profile by  
changing direction of the neutrino trajectory.  
This happens, for the solar neutrinos 
for fixed position of a detector due to the Earth rotation. 
Let $\eta$ be the nadir angle of neutrino trajectory and  
$\eta_s$ corresponds to the border of the $s-$layer,  
so that for  $\eta <  \eta_s$, 
neutrino crosses both the $d-$ and $s-$layers,   
whereas for $\eta > \eta_s$  it crosses the $d-$ layer only. 
The length of trajectory in the layers depends on 
$\eta$.  If $x_d > \lambda_{att}$, the oscillations are averaged in 
the $d-$layer and therefore for $\eta > \eta_s$    
the probability $P^0_e$ does not depend on 
$\eta$. For $\eta < \eta_s$ one can observe 
the oscillatory dependence of the probability 
$P_e$ on $\eta$ induced by the structure,  
since the length of trajectory in the $s-$layer, $x_s$, changes with $\eta$. 
Therefore in what follows we will quantify the 
effect of $s-$layer by the depth of this oscillatory pattern  
$$
D_e \equiv |\Delta P^{max}_e| = |P^{max}_e - P^{min}_e|,
$$ 
and by difference of the averaged probabilities with,  
$\bar{P}_e$, and without, $\bar{P}_e^0$,  the structure: 
$$
\Delta \bar{P}_e \equiv \bar{P}_e - P^0_e.  
$$
In many cases $P^{max}_e = P^0_e$ and $\Delta \bar{P}_e = 0.5 D_e$. 
Apart from $D_e$ and  $\Delta \bar{P}_e$ 
information about  the structure is contained 
in the period and phase of the oscillatory pattern at 
$\eta < \eta_s$. 

We will  describe the attenuation effect in terms of the 
diameter of precession in the $s-$layer, $D_s$,   
and its projection onto one of the eigenstates axes involved. 
Projection of $D_s$  onto the  flavor 
axis is determined  by the flavor mixing angle and therefore does 
not produce smallness. \\

Depending on type of density profile and channel of oscillations 
we obtain the following  results.

1. Let us consider the $\nu_1 \rightarrow \nu_e$ 
transition in the  profile with a remote structure. 
Evolution of the neutrino vector is shown  
in Fig.~\ref{fig:f1c} left.  
The initial state is ${\bf P}(0) = 0.5 {\bf A}_v$. 
In the $s-$layer it precesses around ${\bf A}_s$  with the cone 
angle $2\theta_s'$, which is  the angle between ${\bf A}_v$ and ${\bf A}_s$.  Maximal 
effect in the $s-$layer  corresponds to 
the state ${\bf P}_1$ or the precession phase  $\pi + 2\pi k$  
($k$ is integer) at the moment   when neutrino  arrives at the $d-$layer:  
$$
s-layer:~~~ {\bf P}(0) \rightarrow {\bf P}(x_s) = {\bf P}_1.   
$$
According to Fig.~\ref{fig:f1c}   
the angle between ${\bf P}_1$ and ${\bf A}_d$ equals  
$2\theta_x = 2 \theta_s' +  2\Delta \theta_m$. 
Notice that ${\bf P}_1 = {\bf P}_x$ in our  consideration in Sec. \ref{sec:decoh}.

In the $d-$layer the  vector  ${\bf P}$  precesses around ${\bf A}_d$  approaching 
its projection onto ${\bf A}_d$: 
$$
d-layer:~~~{\bf P}_1 \rightarrow  {\bf P}_2 = 
\frac{1}{2} \cos (2 \theta_s' +  2\Delta \theta_m) {\bf A}_d. 
$$
Without the structure we  have 
$$
{\bf P}(0) \rightarrow {\bf P}^0(x_s + x_d) = {\bf P}_1' = \frac{1}{2}\cos 2\theta_d' {\bf A}_d = 
\frac{1}{2} \cos (2 \theta_s' -  2\Delta \theta_m) {\bf A}_d.
$$ 
Projection of the difference of vectors $[{\bf P}_2 - {\bf P}_1']$ 
onto the flavor axis ${\bf z}$  equals 
\be
D_e = - \Delta P(\nu_1 \rightarrow \nu_e )^{max}
= \cos 2 \theta_d \sin 2 \theta_s' J_m 
\label{eq:diffp2}
\ee
giving the depth of the $\nu_e-$oscillations. 
Since $J_m \sim  \sin 2 \theta_s' \sim \epsilon$, we obtain 
$\Delta P(\nu_1 \rightarrow \nu_e )^{max} \sim \epsilon^2$ in accordance to our 
consideration above. Attenuation is realized.  
If oscillations in $s$ are averaged, the effect  of the structure is 
$\Delta \bar{P}_e = 0.5 D_e$.  
Notice that oscillations correspond to change of the vector ${\bf P}$
between positions ${\bf P}_2$ and ${\bf P}_1'$ in the Fig.~\ref{fig:f1c}.

In other terms, the diameter of precession in the $s-$layer equals  
$D_s = \sin 2 \theta_s'$. Its projection onto 
${\bf A}_d$ (forced by the  averaging) is 
$D \sin 2 \Delta \theta_m = \sin 2 \theta_s' J_m$, 
and finally projection onto the flavor axis is given    
$D_e  = \cos2\theta_d \sin 2\theta_s' J_m$. 
Thus, the origins of the attenuation  are  (i) smallness of  
the diameter of precession $D_s \sim \epsilon$,  
which  is due to the initial mass state $\nu_1$,  
(ii) projection of $D_s$ onto the 
eigenstate axis  ${\bf A}_d$ forced by the averaging in $d$,  
this produces another smallness $\epsilon$.

Without structure after complete averaging in the $d-$layer we have 
\be 
\bar{P}^0_e(\nu_1 \rightarrow \nu_e)  =    \bar{P}^0_1(\nu_e \rightarrow \nu_1)  =
\frac{1}{2}(1 +  \cos 2\theta_d \cos 2\theta_d'). 
\label{eq:without-s}
\ee
Without averaging maximal value of $P_{e}(\eta)$ 
equals $\cos^2 \theta$.

Performing scanning of the profile we will observe at $\eta > \eta_s$ 
the constant probability $\bar{P}^0_e$ of (\ref{eq:without-s}). For $\eta < \eta_s$ 
the probability oscillates around the average value $\bar{P}_e = \bar{P}^0_e - 0.5D_e$
with the depth $D_e \sim \epsilon^2$ (\ref{eq:diffp2}). So that  $P_{e}^{max}(\eta) = \bar{P}^0$. \\

2. Let us consider the $\nu_1 \rightarrow \nu_e$ transition in 
the matter profile with structure near a detector  
(see  Fig.~\ref{fig:f1c} right).  
In the  $d-$layer  the initial state ${\bf P}(0) = 0.5 {\bf A}_v$ 
evolves to its averaged value:  
\be
d-layer:~~~{\bf P}(0) \rightarrow {\bf P}_1  
= \frac{1}{2}\cos 2 \theta_d' {\bf A}_d. 
\label{eq:inst}
\ee
Without structure this gives final position ${\bf P}^0 = {\bf P}_1'$. 
Then in the $s-$layer the vector  
${\bf P}$ precesses around ${\bf A}_s$ with the cone  angle 
$2\Delta \theta_m$, see Fig.~\ref{fig:f1c} right. 
The diameter  of precession equals  
2$|{\bf P}(x_d)| \sin 2\Delta \theta_m$. Its projection onto the  flavor axis  gives 
the depth of $\nu_e-$oscillations due to the structure: 
\be
D_e = - \Delta P(\nu_1 \rightarrow \nu_e )^{max}
= \cos 2 \theta_d' \sin 2 \theta_s J_m.  
\label{eq:diffp1}
\ee
Here $2\theta_s$ is the angle between the axis ${\bf A}_s$ and ${\bf z}$.   
$\theta_s \approx \theta_{12}$  is 
the flavor mixing angle in the layer $s$, and it is large.   
According to (\ref{eq:diffp1}) 
$D_e \approx  J_m \sin 2\theta_{12} \sim \epsilon$,  the 
effect of structure  appears in the  lowest order in $\epsilon$, {\it i.e.} 
the  attenuation is absent. Change of the averaged probability 
equals $\Delta \bar{P}_e = - 0.5D_e$.  

In contrast to the first case here the  projection of ${\bf P}$ onto 
${\bf A}_d$ (induced by averaging) occurs {\it before} the oscillations 
in the $s-$layer.
Averaging changes the diameter of the precession in $s$ 
by factor ${\cal O}(1)$, and therefore does 
not produce additional smallness. 
The diameter of precession in $s$: 
$D_s \sim \epsilon$. 
It should be projected onto the flavor axis immediately  which 
does not produce additional smallness. 
Thus, the oscillation effect of the close to  a detector 
structures is not suppressed. 

In this case, for $\eta < \eta_s$ one will observe oscillations with large 
depth $D_e \sim \epsilon$ (\ref{eq:diffp1}) and the average value 
$\bar{P}_e = \bar{P}_e^0 - 0.5D_e$. Furthermore,  
$P_e^{max} = \bar{P}_e^0$ (\ref{eq:without-s}). \\

\begin{figure}[!]
\hspace{0.1cm}
\includegraphics[width=0.4\textwidth, height=0.6\textwidth]{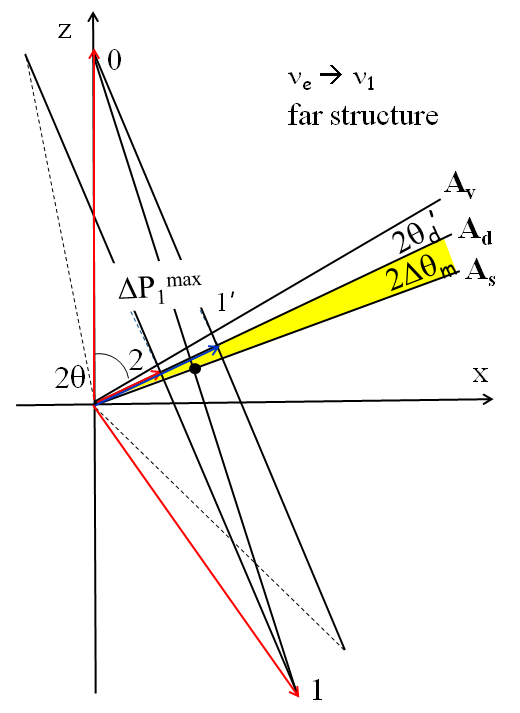}
\includegraphics[width=0.4\textwidth, height=0.6\textwidth]{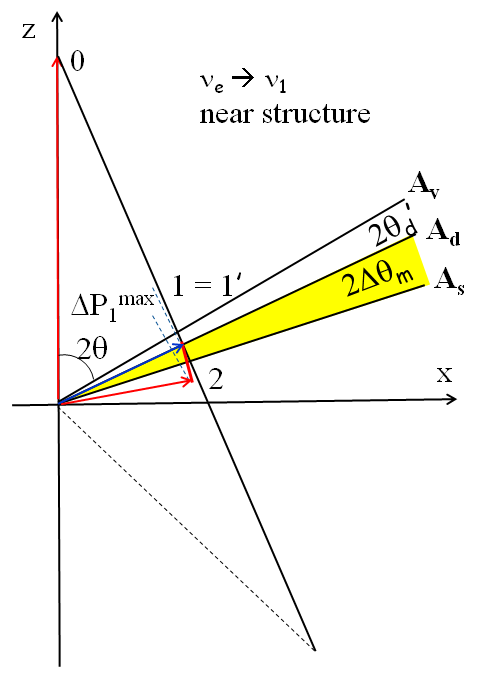}
\caption[�]{The same as in Fig. \ref{fig:f1c}, but  
for  the case of $\nu_e \rightarrow \nu_1$ transition. 
{\it Left panel:} remote structure,  
the attenuation is absent.  {\it Right panel:} 
near structure,     
the attenuation is realized.
\label{fig:f2c}
}
\end{figure}

3. For comparison let us consider the inverse (although not practical) 
case of the  flavor-to-mass, $\nu_e \rightarrow \nu_1$,  transition.  
Now the  initial state, $\nu_e$, is described by  ${\bf P}(0) = 0.5 {\bf z}$.  
The evolution of ${\bf P}$ 
in the matter profile with remote structure is shown in Fig.~\ref{fig:f2c} left.  
In the $s-$layer ${\bf P}$  precesses around ${\bf A}_s$ 
with large ${\cal O}(1)$ diameter  
and evolves to  ${\bf P}_1$  in the case of maximal final effect:  
$$
s-layer: ~~~{\bf P}(0) \rightarrow  {\bf P}(x_s)  = {\bf P}_1.  
$$
${\bf P}_1$ has the angle $(2 \theta_s + 2 \Delta \theta_m)$ with respect to ${\bf A}_d$.  
In the $d-$layer (precessing around ${\bf A}_d$)  ${\bf P}$  converges to 
its projection onto ${\bf A}_d$ (position 2):  
\be
d-layer: ~~~ {\bf P}(x_s) \rightarrow  {\bf P}_2 = \frac{1}{2} 
\cos (2\theta_s + 2 \Delta \theta_m) {\bf A}_d.  
\label{eq:pfin}
\ee 
Without the structure we have 
\be
{\bf P}^0(x_s + x_d) =  {\bf P}_1' = \frac{1}{2} \cos 2\theta_d {\bf A}_d = 
 \frac{1}{2} \cos  (2\theta_s - 2 \Delta \theta_m) {\bf A}_d. 
\label{eq:pfin0}
\ee
Projection of the difference $[{\bf P}_2 - {\bf P}_1']$ 
onto the mass eigenstates axis ${\bf A}_v$ [$({\bf A}_v \cdot{\bf A}_d) = \cos 2\theta_d'
\approx 1$] gives   
the difference of probabilities with and without structure: 
$\Delta P(\nu_e \rightarrow \nu_1) = D_e$ and the latter is given in (\ref{eq:diffp1}).
Thus, 
\be
\Delta P(\nu_e \rightarrow \nu_1)_{far}  =  \Delta P(\nu_1 \rightarrow \nu_e)_{near} 
\label{eq:avdp}
\ee
This coincidence  is  
the consequence of  the $T-$invariance of the physical setup. Namely,    
the shape of the density profile is time-inverted: 
$(s - d) \rightarrow (d - s)$,   and the initial and final states are permuted 
$\nu_1 \leftrightarrow \nu_e$.

Thus,  $\Delta P(\nu_e \rightarrow \nu_1) \approx 
\sin 2\theta_{21} J_m \sim \epsilon  $, {\it i.e.},  
the effect of structure is not 
attenuated in spite of its remote position. Here appearance of 
single factor $\epsilon$ is related to 
projection of the precession diameter, $ {\cal O}(1)$, onto ${\bf A}_d$, 
which is given by  $\sin 2\Delta \theta_m$. 
The observational features are  the same as in the case 2. 
Notice that still averaging leads to suppression: the final depth of oscillations is 
 $ {\cal O}(\epsilon)$  rather than $ {\cal O}(1)$. \\ 

4.  In contrast to the previous case 
the near to detector structure is not visible in  the $\nu_e \rightarrow \nu_1$ transition,  
see Fig.~\ref{fig:f2c} right. Now in the $d-$ layer
\be
d-layer:~~~{\bf P}(0) \rightarrow {\bf P}_1 
= \frac{1}{2}\cos 2 \theta_d {\bf A}_d, 
\label{eq:inst4}
\ee
and large initial precession diameter vanishes. 
In the $s-$layer precession proceeds 
with small angle $2\Delta \theta_m  \sim \epsilon$ around ${\bf A}_s$, 
and then the diameter  of this precession should be 
projected onto  the axis ${\bf A}_v$ 
(given by $\sin 2\theta_s' $)
which leads to  another $\epsilon$. As a result,  we obtain the difference 
of the probabilities with and without structure as in Eq. (\ref{eq:diffp2}). 
Again this is a consequence of the $T-$invariance of the setup. 

The origin of attenuation here is 
(i) reduction of the diameter of precession in $s$: 
$D_s = \cos 2\theta_d J_m \sim \epsilon$  
due to averaging in the $d-$layer,  and  
(ii) projection of $D_s$ onto the mass axis 
${\bf A}_1$ since the final state is $\nu_1$.   
This gives another $\epsilon$.

So, on the contrary to $\nu_1 \rightarrow \nu_e$,  in the 
$\nu_e \rightarrow \nu_1$ channel  
the detector ``sees'' the remote structures,  but 
the closest ones are attenuated. In a sense,  the ``$\nu_1-$detector'' 
is focused on remote structures, when the initial state is $\nu_1$.\\   

In general,   expressions for the depth of oscillations   
induced by  the structure,  $D_e$,   have the form of product 
of the jump factor and the projection factors 
corresponding to initial and final states.  
In four cases considered above  the probabilities are given by two  formulas 
(\ref{eq:diffp2}) with  and  (\ref{eq:diffp1}) without the attenuation.  
Both contain the jump factor. 
They differ by the projection factors in which the flavor and mass mixing angle are 
permuted: $\theta_d  \leftrightarrow  \theta_d'$ and 
$\theta_s  \leftrightarrow  \theta_s'$. The attenuation is related to the latter 
-- the mixing of the $s-$layer. Sines of these angles enter the diameter of precession,  
and consequently,  appearance of small mass mixing 
angle $\theta_s'$ gives an additional smallness. \\

Notice that in the cases of attenuation 1 and 4, two small factors  $J_m$ and   
$\sin  2\theta_s'$ play different roles: in the  first case  $\sin  2\theta_s$ determines the 
diameter of precession, whereas  $J_m$ gives the projection onto the  
axis of eigenstates. In the case 4 - {\it vice versa}. \\ 

5. Finally, let us consider a remote structure and the 
$\nu_e \rightarrow \nu_e$ transition. It is  similar to the case  described in 
Fig. \ref{fig:f2c} left,  
but now  the difference of vectors in  
(\ref{eq:pfin}) and (\ref{eq:pfin0}) should be projected onto the flavor axis 
which does not produce a smallness: 
\be
\Delta P(\nu_e \rightarrow \nu_e)^{max}
= - \cos 2 \theta_d \sin  2\theta_s J_m.
\label{eq:avdp1}
\ee
$\cos 2 \theta_d'$ in  (\ref{eq:inst}) is substituted  here by  
$\cos 2 \theta_d$. 
So,  both projections are given by large flavor mixings and 
there is no attenuation as in the case 3. 

For a near structure (and $\nu_e \rightarrow \nu_e$ mode) we obtain the same result as in 
(\ref{eq:avdp1}) due to the $T-$invariance. \\

Let us consider  generalization of the formalism to the case when density 
in the $d-$layer changes adiabatically. (The same can be done for the $s-$layer). 
Now  the jump factor $J_m$ is determined by 
difference of the densities immediately before a  jump and after a jump. 
The oscillation phase should be computed by integration  (\ref{phasexl}).   
The angles $\theta_d$ and  $\theta_d'$ in the  projection factors 
should be taken at the outer border the layer $d$ which is not attached to the $s-$ layer. 
Thus, in the case 1 the result is given by eq. (\ref{eq:diffp2}) with substitution 
$\theta_d \rightarrow \theta_d^f$ in the projection factor,  
where $\theta_d^f$ is the mixing angle at the end of the layer $d$ 
({\it i.e.},  near a detector). 
In the case 4 the substitution $\theta_d \rightarrow {\theta_d}^{i}$ 
should be done.   
In the second case one should change  
$\theta_d' \rightarrow {\theta_d'}^{i}$ in Eq. (\ref{eq:diffp1}), where 
${\theta_d'}^{i}$ is the mixing angle in matter in the beginning of 
layer $d$.  In the third case:  $\theta_d' \rightarrow {\theta_d'}^{f}$, 
see Eqs. (\ref{eq:avdp}).\\

\section{Attenuation in  multi-layer medium. Two jumps case
\label{sec:threelayers}}

\begin{figure}[!]
\hspace{0.1cm}
\includegraphics[width=0.5\textwidth, height=0.5\textwidth]{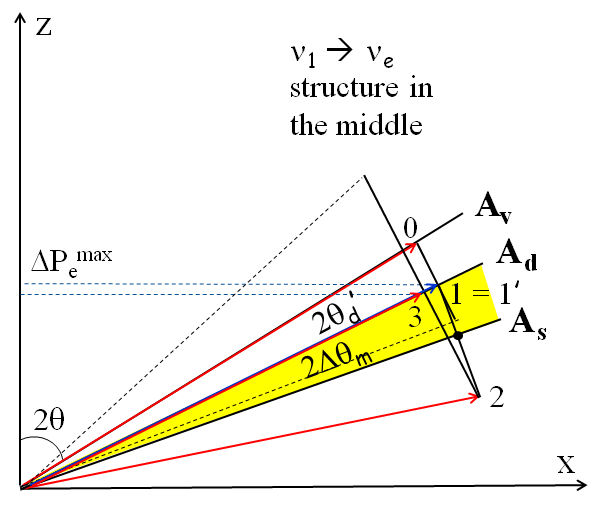}
\caption[�]{The same as in Fig. \ref{fig:f1c}, but 
for  the case of  profile with two jumps 
and the $\nu_1 \rightarrow \nu_e$ transition. 
The diameter of precession $\sim \epsilon$, its projection is given by $\epsilon$. 
The attenuation is present.
\label{fig:f3c}     
}
\end{figure}

Let us consider a matter density profile with three layers: 
$d_1 - s- d_2$, thus adding another decoherence layer to the profile 
studied in Sec.~\ref{sec:twolayers}. Now  there are two jumps.  
We assume that the layers $d_1$ and  $d_2$ have the same properties: 
lengths $x_d$ and densities, and therefore  the same 
eigenstate axis ${\bf A}_d$ 
with the direction fixed by  $2 \theta_d$. The overall profile 
is symmetric with respect to the center 
and similar to the Earth density profile, when  
$d_i$ are identified with the mantle layers,  whereas  $s$ -- with the core.  
The core appears here as the ``structure''.  
Similar matter profile appears also when neutrino crosses two outer shells of the mantle. 
We call it as the profile with structure in the middle.  
As before,  we assume that oscillations in  $d_1$ and  $d_2$ 
are averaged, while in $s$ -- do not.\\   

Let us consider the $\nu_1 \rightarrow  \nu_e$ oscillations 
when $\nu_1$ enters $d_1$,  see Fig. \ref{fig:f3c}. The initial state is  
${\bf P}(0) = 0.5 {\bf A}_v$. 
In the first mantle  layer ${\bf P}$ precesses around ${\bf A}_d$ converging to 
its projection on  ${\bf A}_d$: 
\be
d_1 - layer:~~~ {\bf P}(0) \rightarrow {\bf P}_1 
= \frac{1}{2} \cos 2 \theta_d' {\bf A}_d.  
\label{eq:la-d1}
\ee
In the $s-$layer ${\bf P}$ precesses around ${\bf A}_s$.  
Maximal effect corresponds to the state ${\bf P}_2$ or 
the precession phase at the end of the layer  $\pi + 2\pi k$,  
where $k$ is integer:  
\be
s-layer:~~~{\bf P}(x_d) \rightarrow {\bf P}(x_d + x_s)^{max} =  
{\bf P}_2. 
\label{eq:la-s}
\ee
The angle between ${\bf P}_2$  
and the axis ${\bf A}_d$ is  $4 \Delta \theta_m$, 
$({\bf A}_d \cdot {\bf P}_2) = \cos 4 \Delta \theta_m$.   
Since the length of the precessing vector in $s$ does not change,  
we have  $|{\bf P}_2| =  |{\bf P}(x_d)| = 1/2 \cos 2 \theta_d'$.  
Notice that the length is smaller than $1/2$  
reflecting departure from pure state due to averaging in $d_1$. 
[${\bf P}(x_d + x_s)$ corresponds to ${\bf P}_x$ in our original consideration.]  

In the third layer due to averaging the vector ${\bf P}$ evolves 
to its  projection onto axis ${\bf A}_d$: 
$$
d_2-layer:~~~ {\bf P}(x_d + x_s)^{max} \rightarrow 
{\bf P}_3  =  \frac{1}{2} \cos 2 \theta_d' 
\cos 4 \Delta \theta_m  {\bf A}_d.
$$
Without $s-$layer  we would have ${\bf P}^0(2x_d + x_p) = {\bf P}_1' = {\bf P}_1$ 
(\ref{eq:la-d1}). 
Then projection of the difference  $({\bf P}_3 - {\bf P}_1')$ onto the flavor axis ${\bf z}$ 
gives the depth of oscillations due to the structure:  
\be
D_e = - \Delta P (\nu_1 \rightarrow \nu_e)^{max} = 
\cos 2 \theta_d' \cos 2 \theta_d J_m^2.   
\label{eq:all-av2}
\ee
Again $\Delta P_{e} \approx \cos 2\theta_{12} J_m^2  \propto \epsilon^2$,  
and the attenuation is realized.

The result can be obtained immediately from Fig. \ref{fig:f3c} as follows. 
The length of ${\bf P}$ precessing in the $s-$layer  equals 
$\frac{1}{2} \cos 2 \theta_d'$  (\ref{eq:la-d1}). 
Then the diameter of precession 
is $D_s =  \cos 2 \theta_d' \sin 2 \Delta \theta_m$; the projection of this diameter onto 
${\bf A}_d$ (driven by averaging in $d_2$) is given by 
$\cos 2 \theta_d' \sin^2 2 \Delta \theta_m$; 
finally,  its projection onto the flavor axis leads to  (\ref{eq:all-av2}).
The change of the average probability (due to structure) equals  
$\Delta \bar{P}_e = 0.5 D_e $.

The origin of attenuation is similar to that in  the case 1 of Sec.~\ref{sec:twolayers}.  
One $\epsilon$ appears because of smallness 
of the precession diameter in the $s-$layer. 
Averaging in $d_1$ gives only small reduction of this diameter.  
Another $\epsilon$ is a result of projection of this diameter 
onto axis ${\bf A}_d$ of the layer $d_2$. This is described by  $\sin 2 \Delta \theta_m$. 
Both precession diameter and the projection onto 
the axis of eigenstates ${\bf A}_d$ are determined by 
$J_m$.

This attenuation is realized in the Super-Kamiokande detection of the solar neutrinos. 
No change of the probability should be seen at $\eta = \eta_s$ in the lowest order in $\epsilon$. 
In the next order ($\epsilon^2$) for $\eta < \eta_s$ one expects oscillations 
below $P^{max}_e = \bar{P}^0$ with depth (\ref{eq:all-av2}) that describes spikes. 
Due to unitarity the decrease of $P_{1e}$ corresponds to the increase of $P_{2e}$. 
Since for high energy part of the solar neutrino spectrum neutrinos 
arrive mainly in the $\nu_2$ mass state, 
the decrease of $P_{1e}$ means increase of the $\nu_e$ signal. \\

\begin{figure}[!]
\hspace{0.1cm}
\includegraphics[width=0.45\textwidth, height=0.6\textwidth]{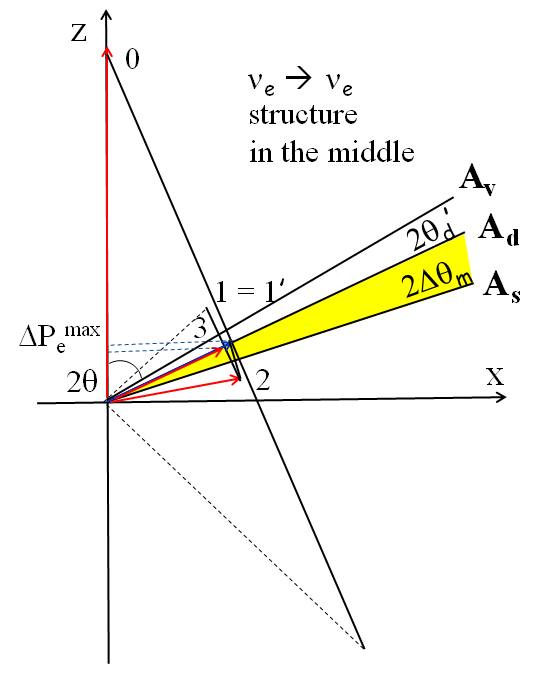}
\caption[�]{The   same as in Fig. \ref{fig:f3c}, 
but for  the $\nu_e \rightarrow \nu_e$ transition. 
The diameter of precession is 
$\sim \epsilon$, its projection is given by $\epsilon$. The attenuation is present.
\label{fig:f4c}
}
\end{figure}

Let us consider the  attenuation  for the flavor channel $\nu_e \rightarrow \nu_e$
(see Fig. \ref{fig:f4c}).  
The result can be obtained immediately from 
(\ref{eq:all-av2}). The only difference is that in the first mantle layer the initial (flavor) 
state is  ${\bf P}(0) = 0.5 {\bf z}$. It evolves to 
$$ 
{\bf P}(0) \rightarrow  {\bf P}_1 =  \frac{1}{2} \cos 2 \theta_d {\bf A}_d,   
$$ 
so that  $\theta_d'$  in Eqs. (\ref{eq:la-d1}) and  (\ref{eq:all-av2}) 
should be substituted by  $\theta_d$.  
Therefore, the final difference of the probabilities with and without core  equals 
\be
D_e = - \Delta P(\nu_e \rightarrow \nu_e)^{max} = \cos^2 2 \theta_d J_m^2.
\label{eq:all-avee}
\ee
$\Delta P(\nu_e \rightarrow \nu_e)^{max}  \sim \epsilon^2$,  
the structure  is attenuated, in contrast to the case of  
$\nu_e \rightarrow \nu_e$ transition 
in $2-$layers of Sec. \ref{sec:decoh}, when suppression was $\sim \epsilon$.
The reason for such a difference is that now the state  which arrives at the structure (core) 
is close to the mass state and therefore the oscillation effect in two other layers 
is similar to that for $\nu_1 \rightarrow \nu_e$ of Sec.~\ref{sec:twolayers}. 
So,  in the case of complete averaging in $d_1$  
the attenuation appears for any initial neutrino state. 
Here averaging in $d_1$ plays crucial role:   
the first layer prepares the incoherent system of states close to the mass eigenstates. 

In both 3 layer cases the factor $J_m$ appears being squared,  which corresponds to the presence of 
two jumps. The projection factors are given by cosines of the mixing 
angles and do not produce additional smallness. 
Now the sign of the effect is fixed and does not depend on the sign of difference of 
densities. 
The observational consequences are as in the previous case: for 
$\eta < \eta_s$  one expects oscillations with the depth (\ref{eq:all-avee}) below 
$\bar{P}_e^0$. 

Notice that effect of  a structure with more than 2 jumps will be still 
proportional to $J_m^2 \sim \epsilon^2$, although some additional numerical factor 
can appear.
Thus,  for 4 jumps of the same size one can find that 
the maximal total effect is proportional to $\sin 4 \Delta \theta_m 
\approx 4 J_m^2$, {\it i.e.}, 4 times larger than in the 2 jumps case
due to the parametric enhancement.

When the density in $d$ layer changes adiabatically
both $\theta_d$ and   $\theta_d'$ 
in Eq. (\ref{eq:all-av2}) should be substituted by their values at the surface.

\section{Attenuation in the case of partial decoherence
\label{sec:partial}}
%

\begin{figure}[!]
\hspace{0.1cm}
\includegraphics[width=0.45\textwidth, height=0.45\textwidth]{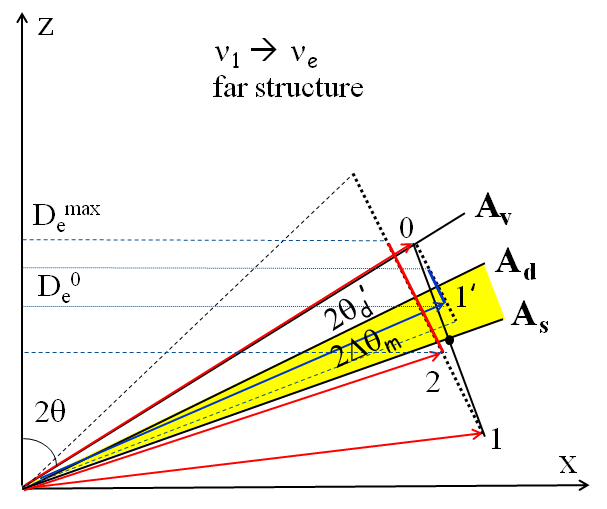}
\includegraphics[width=0.45\textwidth, height=0.45\textwidth]{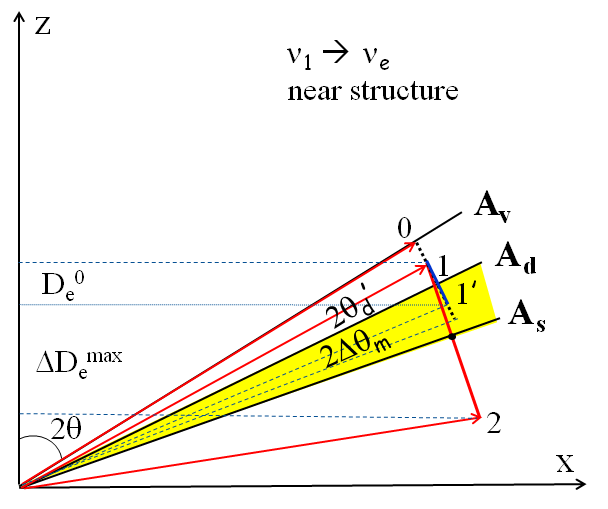}
\caption[�]{The   same as in Fig. \ref{fig:f1c} 
(remote structure, $\nu_1 \rightarrow \nu_e$ transition), but with partial averaging 
of oscillations in the $d-$layer. Red and blue sections 
show final precession diameters in the cases of profiles with and without structure correspondingly.  
{\it The left panel:} remote structure.    
{\it The right panel:}  near structure.
The attenuation is absent in both cases.
\label{fig:f5c}
}
\end{figure}

Partial averaging (decoherence)  of oscillations can be described by the 
factor $\xi \leq 1$ in the interference term of oscillation probability  
which suppresses  the depth of oscillations. 
For $\xi \neq 0$ the effect of structure also depends on the phase of precession in the $d$ layer, 
and we will 
find maximal possible effect  of the structure varying  this phase.  
It can be shown that in the graphic representation the same 
parameter $\xi$ describes reduction of the precession diameter:  
$D \rightarrow \xi D$, projection of ${\bf P}$ onto the precession axis does not change. 
Notice that $\xi$ as function of $x$ depends on the shape of wave packets. 
For wave packets with the exponential tails we have $\xi = exp(- s/\sigma_x)$, where 
$s\approx  (\Delta m^2/2E^2) t$ is the relative shift of the packets due 
to difference of the group velocities 
and $\sigma_x$ is the width of packet. In this case $\xi$ obeys multiplicative properties: 
if $\xi_1$ and $\xi_2$ are the averaging factors in the layers 1 and 2,  
the total averaging of oscillations after crossing  both layers 
is given by the product $\xi^{tot} = \xi_1 \xi_2$. 

In what follows we will consider effects of $\xi \neq 0$ for some cases presented in the 
previous sections. 
 
1. Let us first study the $\nu_1 \rightarrow \nu_e$ oscillations in the   
2 layers profile with remote structure (see Fig.~\ref{fig:f5c} left).  
In the $s-$layer oscillations proceed  as the case 1 of Sec.~\ref{sec:decoh}.    
The phase $\phi_s$ acquired in this layer determines 
characteristics of oscillations in the layer $d$,  in contrast to 
complete averaging case. 
Maximal effect of oscillations corresponds to the state ${\bf P}_1$ 
at the end of $s$ when the  phase equals $\phi_s = \pi + 2\pi k$.   
Then precession of  ${\bf P}$ in the $d-$layer around the axis ${\bf A}_d$  
will be with the initial diameter  $D_d = \sin (2\theta_s' + 2\Delta \theta_m)$.  
After partial averaging in $d$ the projection of $D_d$  onto the flavor axis equals 
\be
D_e^{max}(x_d)  = \xi  \sin (2\theta_s' + 2\Delta \theta_m) \sin 2 \theta_d 
= \xi  \sin (2\theta_d' + 4\Delta \theta_m) \sin 2 \theta_d. 
\label{eq:depth1} 
\ee
Here $\xi = \xi(x_d)$. Thus,  $D_e^{max} \sim \epsilon$.  

Without the structure  the diameter  of precession in the beginning would be 
$\sin 2\theta_d'$. Partial averaging reduces it down to $\xi \sin 2\theta_d'$,  
and projection of the diameter onto the flavor axis gives   
\be
D_e^{0}  = \xi  \sin 2\theta_d' \sin  2 \theta_d.
\label{eq:depth}
\ee 
So, in the absence of structure  
the probability $P_e^0$ oscillates with 
the depth $D_e^0(x_d)$ (\ref{eq:depth}) around average 
value given in Eq. (\ref{eq:without-s}).
The oscillation depth decreases with increase of $x_d$. 
The probability  oscillates below maximal value $P_e = P^{max}_e \approx \cos^2 \theta$. 
 
In the presence of structure, the probability $P_e$  
oscillates around nearly the same average value as without the structure  
(the same in the lowest approximation in $\epsilon$),  but with bigger depth and 
the maximal possible depth is given in (\ref{eq:depth1}). 
Now $P_e^{max}$ can be even above $P = \cos^2 \theta$,   
which  is a manifestation of the parametric enhancement of oscillations. 
This type of the oscillation pattern has been found  in~\cite{Ioannisian:2015qwa}.

The difference of the depths of oscillations with and without 
structure equals   
$$
D_e^{max} -  D_e^{0} = 
\xi [\sin (2\theta_d' + 4\Delta \theta_m) - \sin 2\theta_d'] 
\sin 2\theta_d
=  2\xi \sin 2\theta_d  \cos  2 \theta_s' J_m  \approx 
2 \xi \sin 2\theta_d J_m. 
$$
That is, the effect of structure is of the order $\epsilon \xi$. 
The difference of average  probabilities  is the same as in Eq. (\ref{eq:diffp2}): 
 $\sim \epsilon^2$, and it does not depend on $\xi$. 
Thus, incomplete averaging leads to difference of depths of precession,
but does not change averaged 
values in the lowest order. This is a consequence of the fact that before 
detection the neutrino vector precesses around the same axis in both cases. 

Recall that the depth oscillations in the presence of structure  can be smaller than 
the one without structure  if the density 
in $s$ is smaller than in $d$. \\

2. Let us consider a structure near  detector and  the  $\nu_1 \rightarrow \nu_e$ channel,  
Fig.~\ref{fig:f5c} right. 
In  the $d$ layer the polarization vector 
precesses around ${\bf A}_d$ with  the diameter of precession 
at the end of the layer 
$$
D_d = \xi \sin 2 \theta_d'.  
$$
It can be expressed in terms of $\bar{\theta}_d'$ 
--  the angle between ${\bf P} (x_d) = {\bf P}_1$ and ${\bf A}_d$:  
\be
D_d =  2|{\bf P}_1| \sin 2 \bar{\theta}_d', 
\label{eq:rdrd}
\ee
where the length of ${\bf P}_1$ at the end of layer $d_1$ equals 
\be
 |{\bf P}_1| = \frac{1}{2}~\frac{\cos 2\theta_d'}{\cos 2\bar{\theta}_d'}. 
\label{eq:pdlength}
\ee
The angle  $\bar{\theta}_d'$ is determined by the equality  
\be
\tan 2 \bar{\theta}_d' =  \xi \tan 2 \theta_d'.
\label{eq:angle-tdbar}
\ee
From Eqs. (\ref{eq:rdrd}), (\ref{eq:pdlength}) and (\ref{eq:angle-tdbar})
we obtain the diameter in $d$
\be
D_d = \cos 2\theta_d'  \tan 2 \bar{\theta}_d'.  
\label{eq:depth22}
\ee

The largest final precession depth in the $s-$layer 
is realized the neutrino vector is ${\bf P}_1$ which corresponds to  
the phase  $\phi_d  = 2\pi k$ at the end of layer $d$.  
In this case the angle of precession in $s$ is  
$(2 \bar{\theta}_d' + 2 \Delta \theta_m)$,   
and consequently,  the diameter  of precession in $s$ equals 
$$
D_s =    2|{\bf P}_d|  \sin (2 \bar{\theta}_d' + 2 \Delta \theta_m) = 
\frac{\cos 2\theta_d'}{\cos 2\bar{\theta}_d'}
\sin (2 \bar{\theta}_d' + 2 \Delta \theta_m). 
$$
Its projection on the flavor axis:  
$$
D_e^{max} = \frac{\cos 2\theta_d'}{\cos 2\bar{\theta}_d'}
\sin (2 \bar{\theta}_d' + 2 \Delta \theta_m) \sin 2 \theta_s. 
$$
So,  $D_e^{max} \sim \epsilon$.  
In the limit of complete averaging,  $\bar{\theta}_d' = 0$, 
the above expression coincides with  (\ref{eq:diffp1}). 
Neglecting the high order corrections it can be rewritten as  
$$
D^{max}_e \approx \sin 2 \theta_s \left[\xi  \sin 2 \theta_d' + 
J_m \right].
$$
The average probability equals $\bar{P} \approx \cos^2 \theta_s$.

Without structure the depth of flavor oscillations 
(z-projection of $D_d$ in (\ref{eq:depth22})) would be  
$$
D_e^0 = 
\sin 2\theta_d \cos 2\theta_d' \tan 2 {\bar \theta}_d' 
\approx 
\xi  \sin 2 \theta_d \sin 2 \theta_d'.  
$$
The average value of the probability:  $\bar{P}^0 = \cos^2 \theta_d$. 

The difference of the depths of oscillations with and without structure 
equals 
\be
D_e^{max} - D_e^0  = \frac{\cos 2\theta_d'}{\cos 2\bar{\theta}_d'}
\left[\sin 2 \theta_s  \sin (2 \bar{\theta}_d' + 2 \Delta \theta_m)  
- \sin 2\theta_d \sin 2 {\bar \theta}_d' \right] 
\approx  2\sin 2 \theta_s J_m,   
\label{eq:diff22}
\ee
which does not depend on $\xi$. So, in the lowest order,  partial averaging affects 
$D_e$ and  $D_e^0$ equally.  The dependence of $(D_e^{max} - D_e^0)$ 
on $\xi$ appears in the next order in $\epsilon$  
being  $\approx 2 \bar{\theta}_d' J_m \sim \xi \epsilon^2$. 
If $\bar{\theta}_d' = 0$ (complete averaging),  we would get from 
(\ref{eq:diff22}) the value  $\Delta P_e = - (D_e^{max} - D_e^0)$,  which coincides with 
that in (\ref{eq:diffp1}).

Difference of the averaged probabilities with and without
structure is large: 
$$
\bar{P} -  \bar{P}^0 \approx - \sin 2 \theta_d J_m \sim \epsilon. 
$$ 
It was no attenuation  even 
in the case of complete averaging in $d$, so that the effect of $s-$layer 
appeared at the level of  $\epsilon$. 

If $\phi(x_d) = \pi k$, then the  maxima of survival probability with and without 
the structure are approximately equal:   
$P^{max}_e (x_d) \approx P^{max}_e(x_d + x_p)$, otherwise the probability 
with structure is smaller than that without it. 
Observational effect consists of  increase of oscillation 
depth and decrease of the average 
probability at $\eta < \eta_s$.\\ 

\begin{figure}[!]
\hspace{0.1cm}
\includegraphics[width=0.45\textwidth, height=0.6\textwidth]{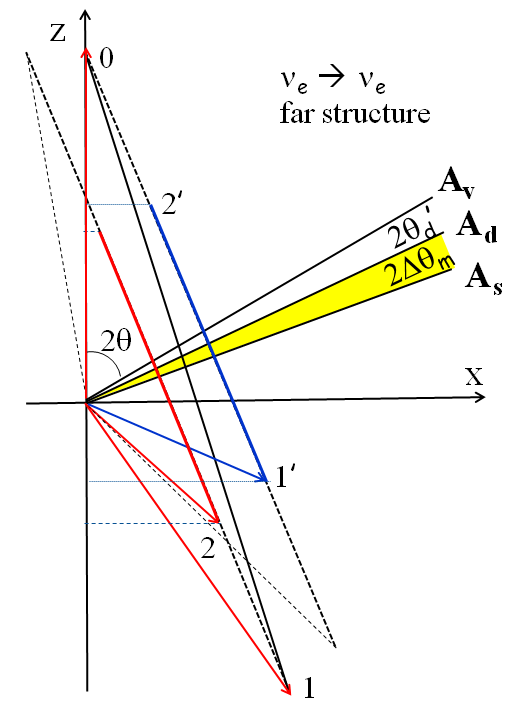}
\caption[�]{The   same as in Fig.~\ref{fig:f5c}
(remote structure) but for                    
the $\nu_e \rightarrow \nu_e$ transition. 
The attenuation is absent.
\label{fig:f6c}
}
\end{figure}

3. Let us consider the $\nu_e \rightarrow \nu_e$ transition 
and remote structure (Fig.~\ref{fig:f6c}). 
Without $s-$layer the diameter of precession in the $d-$layer equals  
$$
D^0 =  \xi \sin 2 \theta_d, 
$$
and its projection onto the flavor axis is  
\be 
D^0_e = \xi  \sin^2 2 \theta_d.
\label{eq:diff-ee1}
\ee
In the limit $\xi = 1$ it coincides with standard oscillation depth.  

With  structure,  precession in the $s-$layer  has the diameter  
$\sin 2 \theta_s$. Maximal final depth of oscillations 
in the $d-$layer corresponds to the vector ${\bf P}_1$  and  the phase 
$\phi(x_s + x_d) = \pi + 2\pi k$.   
The angle between ${\bf P}_1$  and  
${\bf A}_d$ is $(2 \theta_s + 2 \Delta \theta_m)$, so that the precession 
diameter  in the beginning of $d-$layer equals 
$$
D^{max}_d =  \sin (2 \theta_s + 2 \Delta \theta_m). 
$$
Taking into account averaging and projecting the diameter  onto the flavor axis 
we obtain the depth of oscillations at a  detector 
\be
D^{max}_e = \xi \sin 2 \theta_d \sin (2 \theta_s + 2 \Delta \theta_m).   
\label{eq:diff-ee2}
\ee
The difference of the depths with and without structure equals 
according to (\ref{eq:diff-ee1}) and (\ref{eq:diff-ee2})
\be 
D_e^{max} - D_e^0 = 2\xi \sin 2\theta_d \cos 2\theta_s J_m. 
\label{eq:avav33}
\ee
The difference is  the order of $\epsilon$, {\it i.e.},  attenuation is absent as 
in the case 5 of  Sec. \ref{sec:decoh}. 

The position of neutrino vector ${\bf P}_2$ corresponds  
to the minimal value of the probability
$P_e^{min}$ at the end. The maximal value of the probability 
in the presence of structure,  $P_e^{max}$, is in the position ${\bf P}_2'$,
which is realized when the phase of precession in
$s$  is $2\pi k$, that is,
the neutrino enters the $d-$layer as ${\bf P}(0)$.
The difference of the maximal and minimal probabilities 
can be found from the  Fig.~\ref{fig:f6c}:
\be
P_e^{max} - P_e^{min} = (\xi \sin 2\theta_d \cos 2\theta_s +  
\cos 2 \theta_d \sin 2 \theta_s) J_m
\label{eq:avavmax}
\ee
which differs from (\ref{eq:avav33}). 
In the limit $\xi \rightarrow  0$ it is reduced to
the expression (\ref{eq:avdp1})  for the complete averaging.

It is straightforward to show that the difference of the average 
oscillation probabilities is the same as in the case 
of complete averaging in the layer $d$, see  (\ref{eq:avdp}). 
So,  here we have oscillations with ${\cal O}(1)$ depth. 
The differences of depths of oscillations and average values 
(with and without structure) are  of the order $\epsilon$. 

Notice that it was no attenuation even with complete averaging. 
Incomplete averaging 
does not change the difference of average probabilities, 
but produces  difference of depths of oscillations 
(\ref{eq:avav33}) of the order $\epsilon$.\\

\begin{figure}[!]
\hspace{0.1cm}
\includegraphics[width=0.5\textwidth, height=0.5\textwidth]{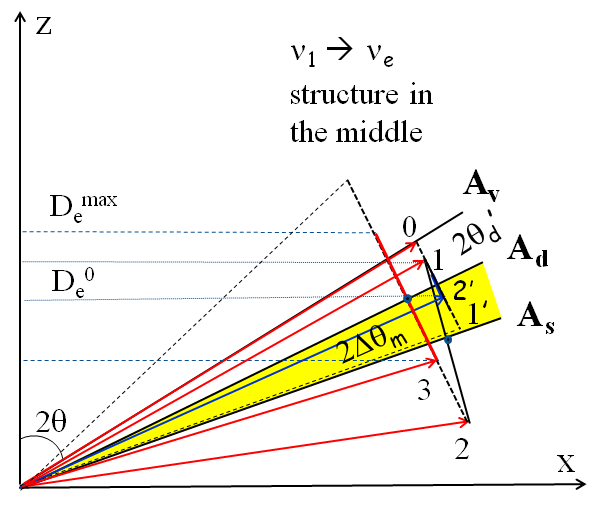}
\caption[�]{The same as in Fig. \ref{fig:f3c},  but with partial averaging in $d_1$  and $d_2$.  
The attenuation is absent.
\label{fig:f7c}
}
\end{figure}

4. Let us consider $\nu_1 \rightarrow \nu_e$ 
transition in the  symmetric 
profile with three layers $(d_1 - s - d_2)$ 
(see Fig. \ref{fig:f7c}).  
If $\xi_1$ and $\xi_2$ are the averaging factor in the  layers $d_1$  and $d_2$,  we obtain 
the depth of oscillations without structure $(d_1 - d_2)$: 
\be
D_e^{max} (x_d) = \xi_1 \xi_2 \sin 2\theta_d' \sin  2 \theta_d,  
\label{eq:depth2}
\ee
which differs from (\ref{eq:depth}) by additional power of $\xi$. 
The average value of probability is  given in (\ref{eq:without-s}). 
    
As in the  case 2 of this section,  we use the angle $\bar{\theta}_d'$ (\ref{eq:angle-tdbar}) 
between the polarization vector at the end of layer $d_1$,  ${\bf P}_1$,  and the axis 
${\bf A}_d$. 
Then the precession angle in the  $s-$layer is $(2 \bar{\theta}_d' + 2 \Delta \theta_m)$. 
The length of ${\bf P}_1$ is given in (\ref{eq:pdlength}).  
Maximal final precession depths corresponds to the phase of oscillations 
in the $s-$layer $\phi_s = \pi + 2\pi k$ ($k-$ integer) 
when neutrino state is described by ${\bf P}_2$. 
The angle of precession in the 
layer $d_2$ -- the angle between ${\bf P}_2$ and  ${\bf A}_d$ 
is  $(2 \bar{\theta}_d' + 4 \Delta \theta_m)$. 
Consequently, the initial diameter of precession in $d_2$ equals 
$$
D = \frac{\cos 2\theta_d'}{\cos 2\bar{\theta}_d'} 
\sin (2 \bar{\theta}_d' + 4 \Delta \theta_m). 
$$ 
Averaging in the layer  $d_2$ gives another factor $\xi_2$,  and then 
projection on the flavor axis leads to  
$$
D_e^{max} = \xi_2 \frac{\cos 2\theta_d'}{\cos 2\bar{\theta}_d'} 
\sin (2 \bar{\theta}_d' + 4 \Delta \theta_m) \sin 2\theta_d.
$$

The oscillations proceed around the average value 
$$
\bar{P}_e = \frac{1}{2}  +  \frac{\cos 2\theta_d'}{2\cos 2\bar{\theta}_d'}
\cos (2 \bar{\theta}_d' + 4 \Delta \theta_m) \cos 2\theta_d.
$$ 
The difference of average probabilities with and without structure  is 
very small $\sim \epsilon^2$.

Difference of the oscillation depth with and without structure equals   
\be
D_e^{max} - D_e^0 = \xi_2 \sin 2\theta_d
\left[\frac{\cos 2\theta_d'}{\cos 2\bar{\theta}_d'}
\sin (2 \bar{\theta}_d' + 4 \Delta \theta_m) -  \xi_1 \sin 2\theta_d'
\right].
\label{eq:difdep}
\ee
(In the limit $\Delta \theta_m = 0$, the expression in the brackets vanishes, as it should be.)
In the lowest approximation in $\epsilon$ Eq. (\ref{eq:difdep}) reduces to  
$$
D_e^{max} - D_e^0  \approx  2 \xi_2 \sin 2\theta_d J_m = O(\xi  \epsilon).   
$$
Thus, the difference of depths is proportional to $\xi_2$,  
whereas dependence on $\xi_1$ is absent. There is no attenuation:  
$D_e^{max} - D_e^0 \propto \xi_2 \epsilon$. 
The difference of averaged probabilities,  
$\Delta \bar{P}_e \sim \epsilon^2$, is slightly changed from that 
in Fig.~\ref{fig:f3c}. 
There is no  change of average values of the probabilities 
in the lowest order in $\epsilon$,  
since oscillations in the last layer occur in both cases (with and without structure)
around the same axis. \\  

\begin{figure}[!]
\hspace{0.1cm}
\includegraphics[width=0.45\textwidth, height=0.65\textwidth]{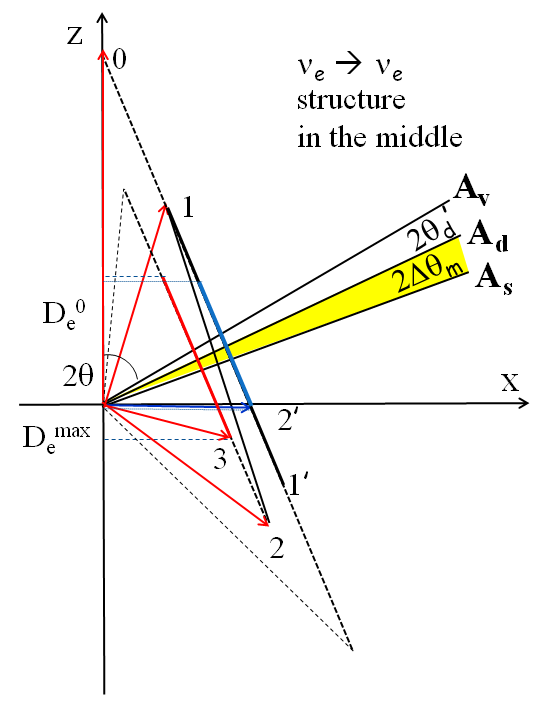}
\caption[�]{The   same as in Fig. \ref{fig:f7c}, 
but for $\nu_e \rightarrow \nu_e$ transition. 
The attenuation is absent.
\label{fig:f8c}
}
\end{figure}


5. Let us  consider the $\nu_e \rightarrow \nu_e$ transition 
in the case of partial averaging 
in the symmetric profile with three layers, Fig. \ref{fig:f8c}. 
The only difference from the previous case is the initial state given now by 
${\bf P}(0) = \frac{1}{2} {\bf z}$,  
instead of  ${\bf P}(0) = \frac{1}{2} {\bf A}_v$. 
Therefore in formulas obtained for the 
$\nu_1 \rightarrow \nu_e$ transition $\theta_d'$ should be substituted by $\theta_d$. 
As in the  case 2,  we introduce  the angle $\bar{\theta}_d$ 
between the polarization vector ${\bf P}_1$ at the end of layer $d_1$ 
after partial averaging 
and the axis ${\bf A}_d$.  It is determined by the equality 
\be
\tan 2 \bar{\theta}_d =  \xi \tan 2 \theta_d.
\label{eq:angle-tdbar2}
\ee
Then the precession angle in the $s-$layer 
(the angle between ${\bf P}_1$ and  ${\bf A}_s$)  is $(2 \bar{\theta}_d + 2 \Delta \theta_m)$.
The length of the vector equals
$$
|{\bf P}_1| =  \frac{\cos 2\theta_d}{2\cos 2\bar{\theta}_d}.
$$
Maximal final depths  of oscillations corresponds to the phases of oscillations
$\phi_d =  2\pi k$ (position ${\bf P}_1$) in the $d_1-$layer and 
$\phi_s = \pi + 2\pi k'$ (position ${\bf P}_2$) in the $s-$layer 
($k$,  $k'-$ are integers) . Under these conditions the parametric enhancement of 
oscillations occur and the precession angle in the $d_2-$layer,  
{\it i.e.} the angle between ${\bf P}_2$ and  ${\bf A}_d$,  
becomes  $2 \bar{\theta}_d + 4 \Delta \theta_m$. 
Consequently, the initial diameter of precession in $d_2$ equals
$$
D^{max} = \frac{\cos 2\theta_d}{\cos 2\bar{\theta}_d}
\sin (2 \bar{\theta}_d + 4 \Delta \theta_m).
$$
Averaging in the layer  $d_2$ gives another factor $\xi_2$,  and then
projection onto the flavor axis leads to
\be
D_e^{max} = \xi_2 \frac{\cos 2\theta_d}{\cos 2\bar{\theta}_d}
\sin (2 \bar{\theta}_d + 4 \Delta \theta_m) \sin 2\theta_d.
\label{eq:withss}
\ee

The depth of oscillations without structure equals
\be
D_e^{max} (x_d) = \xi_1 \xi_2  \sin^2  2 \theta_d, 
\label{eq:depth3}
\ee
which again differs  from (\ref{eq:depth2})  by substitution  
$\theta_d' \rightarrow \theta_d$.  
The difference of the depths with (\ref{eq:withss}) and without 
(\ref{eq:depth3}) structure, 
$$
D_e^{max} - D_e^0 = \xi_2 \sin 2\theta_d
\left[\frac{\cos 2\theta_d}{\cos 2\bar{\theta}_d}
\sin (2 \bar{\theta}_d + 4 \Delta \theta_m) -  \xi_1 \sin 2\theta_d \right], 
$$
is similar to that in (\ref{eq:difdep}) with substitution $\theta_d' \rightarrow \theta_d$ 
in the parenthesis. It can be rewritten as 
\be
D_e^{max} - D_e^0 = \xi_2 \sin 2\theta_d
\left[\cos 2\theta_d \cos 2 \Delta \theta_m J_m - 
\xi_1 \sin 2\theta_d J_m^2 \right] 
\approx   \xi_2 \sin 4\theta_d J_m.
\label{eq:diff-rad}
\ee
There is no attenuation and $D_e^{max} - D_e^0 \propto \xi_2 \epsilon$. 
Attenuation is reproduced if $\xi_2 = 0$, {\it i.e.} in the case of complete averaging in the 
$d_2$ layer, then the diameter of precession becomes zero in the lowest order. 
Dependence on  $\xi_1$ appears in the $\epsilon^2$ order, 
that is,  attenuated. 

The oscillations proceed around the average value
$$
\bar{P}_e = \frac{1}{2}  +  \frac{\cos^2 2\theta_d}{2\cos 2\bar{\theta}_d}
\cos (2 \bar{\theta}_d + 4 \Delta \theta_m).
$$
Without the structure we would have 
$\bar{P}_e^0 = 0.5(1  +  \cos^2 2\theta_d)$. 
The difference of average probabilities with and without structure  
equals 
\be
\Delta \bar{P}_e = - \cos^2 2\theta_d    
\left[ \xi_1  \tan 2\theta_d \cos 2 \Delta \theta_m J_m + J_m^2 \right].                
\label{eq:diff-av}
\ee
Now partial averaging leads to $\Delta D_e  \sim \xi_2 \epsilon$ and 
$\Delta \bar{P}_e \sim \xi_1 \epsilon$. 
Thus,  the difference of average values depends on $\xi_1$ but 
it does not depend on $\xi_2$.  
In the limit $\xi_1 = 0$:   $\Delta \bar{P}_e \propto \epsilon^2$
and the attenuation is recovered. 
Notice that the differences of depth and average values depends on different $\xi$: 
$\xi_2$ and $\xi_1$ correspondingly. This can be used for tomography. \\

The structure in the density profile changes the  depth of precession 
which can be larger or smaller than that without structure 
depending on phases of oscillations in the 
$d_1-$ and $s-$layers.  In the case of smaller density in the $s-$layer than in the 
$d$ layers, the sign of $2 \Delta \theta_m$, and consequently,  the sign of 
$J_m$ change. As  a result,  the difference of precession diameters 
(\ref{eq:diff-rad}) becomes negative 
and the difference of average values (\ref{eq:diff-av}) becomes positive. 

For $\xi \gg \epsilon$,  the effect of structure  appears 
in the lowest order in $\epsilon$ in all the cases (channel, profile),  
although it can be  suppressed by $\xi$.

\section{Discussion and conclusions
\label{sec:discussion}}



1. Attenuation effect is the effect of  loss of sensitivity 
of the oscillation signal  to remote structures 
of density profile due to finite neutrino energy resolution. 
We presented the graphic (geometric) description of the effect. 
We show that the effect is a result of 

\begin{itemize}

\item 
small mixing of the mass states 
in matter;  

\item
incoherence of the neutrino state arriving at a structure; 

\item 
averaging of oscillations (loss of coherence) 
between a structure and a detector. 

\end{itemize}

Contributions to the oscillation effect 
of structures at distances larger than the attenuation length 
are suppressed by additional power of $\epsilon$.  

The attenuation length is 
the distance over which oscillations integrated 
over the  energy resolution interval of neutrinos are averaged. 
In other terms, it is a distance over which the wave packets of the size determined 
by the energy reconstruction function are separated in space. 
The attenuation is realized in the lowest order in $\epsilon$. 
The remote structures produce effects in the $\epsilon^2$ order. 
The better the relative energy resolution $\sigma_E/E$, 
the more remote structures can be seen.

The conditions of  attenuation are valid for a multi-layer medium. 
In the case of several different structures the conditions should be applied 
to each structure independently. 
Interplay between different structures will show up in the next order in 
$\epsilon$. Actually,  we saw this interplay in the case of two jumps.

2. The effect of remote structure is proportional to the change of the  
mixing parameter $J_m \equiv \sin 2 \Delta \theta_m \sim \epsilon$ and the projection factors.  
The attenuation is realized if one of the projection factors is $\sim \epsilon$. 
For the profile with core (two jumps) the jump factor appears as $J_m^2$ in the probability.   
For more than 2 jumps the effect is still proportional to $J_m^2$ with some 
additional coefficients. 

3. In terms of graphic representation the effect of structure  
is determined  by the diameter of precession and its projection onto the 
eigenstate axis (which depends on setup and channel of transition). 
This  allows us to understand immediately why 
in the case of flavor to mass transition  $\nu_e \rightarrow \nu_1$ 
the sensitivity is mainly to remote structures (see \cite{Ioannisian:2004jk}). 
The detector of neutrinos $\nu_1$ is ``focused'' on structures to which neutrino state 
$\nu_1$ arrives.

Graphic description allows us to explicitly compute effects 
in $\epsilon^2$ and higher orders and also obtain results for different  
positions of a   structure and channels of oscillations. 

Attenuation is a result of (i) suppression of the precession diameter  
in the $s-$layer either due to specific initial state (state arriving at the structure)   
or due to averaging, and (ii)  smallness of projection of the diameter 
onto  the eigenstates axis. 

4. In the case of partial averaging the attenuation is absent or weak. 
For all the configurations (channels, profiles) the effect appears 
in the lowest order in $\epsilon$. 
In expressions obtained for complete attenuation 
one factor  $\epsilon$ is substituted by  $\xi$, 
and the effect is given by $\xi \epsilon$. So, it may be suppressed,  
if  $\xi$ is small. 

5. From the observational point of view,  
in the case of complete averaging one 
will see constant $P_e$ at  $\eta > \eta_s$ and 
the oscillatory pattern  at $\eta <  \eta_s$.  
In the case of attenuation the depth of oscillations and change of the 
average probability are of the order $\epsilon^2$.
In absence of the attenuation these parameters are 
of the order  $\epsilon$.

6. Similarly, one can consider the  attenuation in the 1-3 channel. 
There are two features here: the vacuum angle is relatively 
small,  so the eigenstate axes are turned closer to the 
flavor axis. Consequently, in the $2\nu-$case we would get the same formulas as before 
with just substitution $\theta_{12} \rightarrow \theta_{13}$,  
and $\epsilon \rightarrow \epsilon_{13} = EV_e/\Delta m^2_{31}$. 
Low density means here $E < 1$ GeV. 
The $2\nu-$case can be realized  
in the region $(0.2 - 1)$ GeV where  the 1-2 phase is small (evolution is frozen).

7. On practical side, the operations of integration over the energy 
(wave function of a detector) and integration of the evolution equation  can be permuted.  
That is, one can first integrate over the neutrino  energy obtaining wave packets 
and then consider  the flavor evolution, or first compute the flavor evolution 
and then perform the energy integration. 
In the first case it is clear that one can simply neglect effects of remote structures 
in  consideration from the beginning.

The attenuation effect should be taken into account at  
interpretation of experimental data on neutrino oscillations in the Earth and in planning of 
future experiments devoted to the Earth oscillation tomography.

\section*{References}



\begin{thebibliography}{99}

\bibitem{Ioannisian:2004jk} 
  A.~N.~Ioannisian and A.~Y.~Smirnov,
  Phys.\ Rev.\ Lett.\  {\bf 93}, 241801 (2004)
  [hep-ph/0404060].


\bibitem{Renshaw:2013dzu}
  A.~Renshaw {\it et al.} [Super-Kamiokande Collaboration],
  Phys.\ Rev.\ Lett.\  {\bf 112} (2014) no.9,  091805
  [arXiv:1312.5176 [hep-ex]].
  K.~Abe {\it et al.} [Super-Kamiokande Collaboration],
  Phys.\ Rev.\ D {\bf 94}, no. 5, 052010 (2016)
  [arXiv:1606.07538 [hep-ex]].

\bibitem{Hosaka:2005um}
  J.~Hosaka {\it et al.} [Super-Kamiokande Collaboration],
  Phys.\ Rev.\ D {\bf 73} (2006) 112001
  [hep-ex/0508053].

\bibitem{Smy:2003jf}
  M.~B.~Smy {\it et al.} [Super-Kamiokande Collaboration],
  Phys.\ Rev.\ D {\bf 69} (2004) 011104
  doi:10.1103/PhysRevD.69.011104
  [hep-ex/0309011].


\bibitem{Ioannisian:2015qwa} 
  A.~N.~Ioannisian, A.~Y.~Smirnov and D.~Wyler,
  Phys.\ Rev.\ D {\bf 92}, no. 1, 013014 (2015)
  [arXiv:1503.02183 [hep-ph]].

  

\bibitem{Ioannisian:2017dkx}
  A.~Ioannisian, A.~Smirnov and D.~Wyler,
  arXiv:1702.06097 [hep-ph].


\bibitem{Akhmedov:2004rq}
  E.~K.~Akhmedov, M.~A.~Tortola and J.~W.~F.~Valle,
  JHEP {\bf 0405} (2004) 057
  [hep-ph/0404083].
  

\bibitem{Maltoni:2015kca}
  M.~Maltoni and A.~Y.~Smirnov,
  Eur.\ Phys.\ J.\ A {\bf 52} (2016) no.4,  87
  [arXiv:1507.05287 [hep-ph]].


\bibitem{Dziewonski:1981xy}
  A.~M.~Dziewonski and D.~L.~Anderson,
  Phys.\ Earth Planet.\ Interiors {\bf 25} (1981) 297.

\bibitem{graphic} 
S. P. Mikheyev and A. Yu. Smirnov, Proc. of the 6th
  Moriond Workshop on massive Neutrinos in Astrophysics and Particle
  Physics, Tignes, Savoie, France Jan. 1986 (eds. O. Fackler and
  J. Tran Thanh Van) p. 355 (1986).
  J.~Bouchez, M.~Cribier, J.~Rich, M.~Spiro, D.~Vignaud and W.~Hampel,
  Z.\ Phys.\ C {\bf 32} (1986) 499.
  P.~I.~Krastev and A.~Y.~Smirnov,
  Phys.\ Lett.\ B {\bf 226} (1989) 341.
  Q.~Y.~Liu, S.~P.~Mikheyev and A.~Y.~Smirnov,
  Phys.\ Lett.\ B {\bf 440} (1998) 319
  [hep-ph/9803415].


\bibitem{Nicolaidis:1990jm}
  A.~Nicolaidis, M.~Jannane and A.~Tarantola,
  J.\ Geophys.\ Res.\ Solid Earth {\bf 96} (1991) 21811.


\bibitem{Lindner:2002wm}
  M.~Lindner, T.~Ohlsson, R.~Tomas and W.~Winter,
  Astropart.\ Phys.\  {\bf 19} (2003) 755
  [hep-ph/0207238].

\bibitem{Akhmedov:2005yt}
  E.~K.~Akhmedov, M.~A.~Tortola and J.~W.~F.~Valle,
  JHEP {\bf 0506} (2005) 053
  [hep-ph/0502154].

\bibitem{Winter:2006vg}
  W.~Winter,
Earth Moon Planets {\bf 99} (2006) 285
  [physics/0602049].


\bibitem{Koike:2016jrb}
  M.~Koike, T.~Ota, M.~Saito and J.~Sato,
  Phys.\ Lett.\ B {\bf 759} (2016) 266
  [arXiv:1603.09172 [hep-ph]].







  







\end{thebibliography}
\end{document}